\documentclass[conference]{IEEEtran}
\IEEEoverridecommandlockouts
\usepackage{cite}
\usepackage{amsmath,amssymb,amsfonts}
 \usepackage{verbatim}
\usepackage{graphicx}
 \usepackage{url} 
\usepackage{mdframed}
\usepackage{multirow}
\usepackage{booktabs}
\usepackage[table,xcdraw]{xcolor}
\usepackage{textcomp}
\usepackage{xcolor}
\usepackage{algpseudocode}
\usepackage{algorithm}
\usepackage[table]{xcolor} 
\usepackage{multirow}      
\usepackage{booktabs}      
\usepackage[utf8]{inputenc}
\usepackage{tcolorbox}
\tcbuselibrary{skins}

\def\BibTeX{{\rm B\kern-.05em{\sc i\kern-.025em b}\kern-.08em
    T\kern-.1667em\lower.7ex\hbox{E}\kern-.125emX}}
\begin{document}

\title{Energy-Aware Code Generation with LLMs: Benchmarking Small vs. Large Language Models for Sustainable AI Programming}
\author{
\IEEEauthorblockN{Humza Ashraf\IEEEauthorrefmark{1}, Syed Muhammad Danish\IEEEauthorrefmark{1}, Aris Leivadeas\IEEEauthorrefmark{2}, Yazan Otoum\IEEEauthorrefmark{1}, Zeeshan Sattar\IEEEauthorrefmark{3}} 
\IEEEauthorblockA{\IEEEauthorrefmark{1}Algoma University, Brampton, Canada\\
\IEEEauthorrefmark{2}École de Technologie Supérieure (ÉTS) Montreal, Canada\\
\IEEEauthorrefmark{3}Ericsson Inc., Ottawa, Canada\\
Emails: \{hashraf, syed.danish, otoum\}@algomau.ca,  aris.leivadeas@etsmtl.ca, zeeshan.sattar@ericsson.com}
}

\maketitle

\begin{abstract}


Large Language Models (LLMs) are widely used for code generation. However, commercial models like ChatGPT require significant computing power, which leads to high energy use and carbon emissions. This has raised concerns about their environmental impact. In this study, we evaluate open-source Small Language Models (SLMs) trained explicitly for code generation and compare their performance and energy efficiency against large LLMs and efficient human-written Python code. The goal is to investigate whether SLMs can match the performance of LLMs on certain types of programming problems while producing more energy-efficient code. We evaluate 150 coding problems from LeetCode, evenly distributed across three difficulty levels: easy, medium, and hard. Our comparison includes three small open-source models, StableCode-3B, StarCoderBase-3B, and Qwen2.5-Coder-3B-Instruct, and two large commercial models, GPT-4.0 and DeepSeek-Reasoner. The generated code is evaluated using four key metrics: run-time, memory usage, energy consumption, and correctness. We use human-written solutions as a baseline to assess the quality and efficiency of the model-generated code. Results indicate that LLMs achieve the highest correctness across all difficulty levels, but SLMs are often more energy-efficient when their outputs are correct. In over 52\% of the evaluated problems, SLMs consumed the same or less energy than LLMs.
\end{abstract}


\section{Introduction}

\begin{figure}
    \centering
    \includegraphics[width=\linewidth]{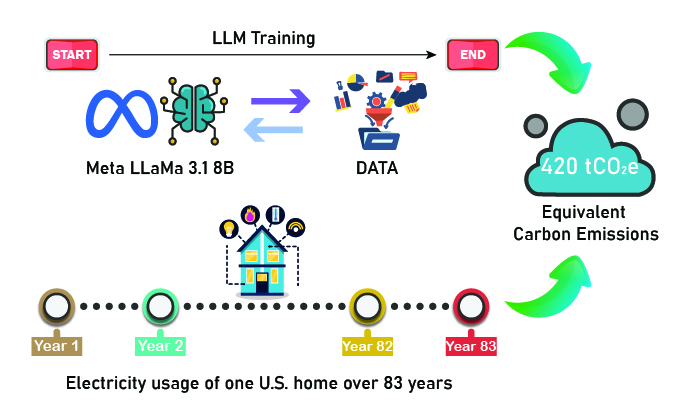}
    \caption{$CO_{2}$ Emissions: LLM Training vs. 83 Years of Home Power \cite{morrison2503holistically}}
    \label{fig:1}
\end{figure}

Large Language Models (LLMs) have achieved tremendous success in Code generation \cite{herrington2003code}; however, there is a growing concern about the impact of software development on the environment. Training and deploying LLMs incurs significant environmental costs, including substantial CO$_{2}$ emissions and water usage \cite{desislavov2023trends}. According to \cite{morrison2503holistically}, training LLaMA 3.1 with 8 billion parameters generated approximately 420 tCO$_{2}$e, which is equivalent to the emissions from 83 years of electricity usage by a single U.S. household, as shown in Fig. \ref{fig:1}. The process also consumed 2,769 kiloliters of water, roughly equal to 24.5 years of water usage by an average American, and this is only for the training phase. Once deployed, these models continue to consume energy as users interact with them. Energy usage during inference has increased rapidly \cite{desislavov2023trends}, and total emissions depend on how frequently the model is used. For example, if ChatGPT receives 100 million queries per day \cite{desislavov2023trends}, and each query consumes about 0.002 kWh of energy, the total daily energy consumption would be approximately 0.2 gigawatt-hours (GWh) \cite{vartziotis2024learn}. As a result of this growing energy demand, LLM-based applications have faced critical questions regarding their sustainability. In light of this, a key question arises: are large LLMs always necessary, especially for routine or simpler coding tasks?

In this context, Small Language Models (SLMs) offer a promising alternative. In general, SLMs use less energy and memory during training and inference due to their simplicity and fewer parameters. For simple or routine tasks, such as solving fundamental coding problems, these models are often faster and more efficient \cite{chen2024role}. Although they are capable of reasoning complex programming challenges, they may be incapable of handling long-term contexts or understanding deep code. Compared to SLMs, LLMs are built with billions of parameters, making them more capable of tackling more difficult coding queries. The improved performance, however, comes at the expense of increased energy demands and greater environmental impact. This opens up a design space for exploring whether smaller, more efficient models can balance performance with sustainability for specific coding tasks.

In this work, we explore that question by conducting a systematic comparison between small and large LLMs in the context of sustainable and efficient code generation. Specifically, we hypothesize that SLMs can match the performance of LLMs on certain types of programming problems and produce more energy-efficient code. Our goal in this study is to identify scenarios where small models can serve as viable, energy-efficient alternatives to large models by evaluating the tradeoffs between performance and resource consumption in code generated by LLMs of various sizes. Notably, we do not measure the energy used to train or run the language models themselves. Instead, our focus is on the energy efficiency of the code produced by these models when executed. We pose the following primary research question (RQ): \textit{How do SLMs and LLMs differ in generating efficient and sustainable code across varying levels of problem complexity?} To answer this \textit{RQ}, this work makes the following key contributions:
\begin{itemize}
    \item We conduct a systematic comparison of three small open-source models, StableCode-3B, StarCoderBase-3B, and Qwen2.5-Coder-3B-Instruct, and two large commercial models, GPT-4.0 and DeepSeek-Reasoner, using standardized code generation prompts. Unlike prior work, our focus is on evaluating whether SLMs can match the code performance of LLMs in terms of correctness and execution efficiency.
    \item We perform extensive experiments on 150 Python programming problems from LeetCode, evenly distributed across easy, medium, and hard categories. For each generated solution, we measure code correctness, energy consumption, memory usage, and runtime in an isolated Linux environment to compare the performance and efficiency of SLMs against LLMs.
\end{itemize}

Results show that LLMs, such as GPT-4.0 and DeepSeek-Reasoner, consistently achieve the highest correctness at all difficulty levels, while exhibiting strong runtime and energy efficiency performance. In contrast, SLMs generally lack accuracy, but models such as Qwen2.5-Coder-3B-Instruct show good generalization and competitive correctness. Notably, when SLMs do produce correct outputs, they are often more energy-efficient, achieving better or equal energy consumption compared to LLMs in over 52\% of the cases. Additionally, findings suggest that not all SLMs perform equally well, and careful model selection is essential, especially in energy-constrained environments.

The rest of the paper is organized as follows. Section \ref{sec:rl} presents the related work. Section \ref{sec:meth} presents the overall methodology. Sec. \ref{sec:result} discusses the analysis and results. Section \ref{sec:ttv} presents the limitations, while Section \ref{sec:con} concludes the paper.

\section{Related Work}
\label{sec:rl}
Many previous studies \cite{du2024mercury, huang2024effibench, yu2024codereval, hendrycks2021measuring, wang2022recode, huang2024effi} have proposed benchmarks to evaluate the code generated by LLMs, but most of them focused on measuring the run-time and memory usage of the generated code. In this study, we also measure energy consumption, giving a more complete view of its efficiency.

Authors in \cite{cappendijk2025exploration} examined how prompt engineering affects the energy consumption of Python code generated by LLMs, aiming to identify strategies for producing more energy-efficient code. Results indicate that specific combinations of prompt modifications lead to reduced energy usage. Authors in \cite{peng2024large} developed a tool for improving the energy efficiency of existing code and to investigate whether LLMs can intelligently refactor code to reduce energy consumption without compromising performance and correctness. Authors in \cite{rani2025can} examined how GPT-3, GPT-4, LLaMA, and Mixtral LLMs can improve energy efficiency in real-world MATLAB code. A total of 400 scripts from 100 popular GitHub repositories are analyzed. LLMs optimize each script, and the optimized code is then evaluated for energy usage, memory usage, execution time, and correctness. Performance is assessed by comparing the LLM-generated code to human-written optimizations. However, this work is different from our primary goal, since we aim to evaluate the energy efficiency of code generated by LLMs, rather than optimize existing code. In addition, our study focuses on the Python programming language and includes SLMs, in contrast to prior work that primarily used large models.

Authors in \cite{cheung2025comparative} examined the environmental impact of LLM-based code assistants compared to human-written code and found that LLM-generated code was computationally more expensive, leading to higher energy consumption. Authors in \cite{coignion2024green} investigated the energy consumption of LLM-based code assistants, such as GitHub Copilot, during software development tasks. Using simulated development sessions based on traces from 20 professional developers, the study examined how factors like model size, quantization, streaming, and concurrency affected energy consumption. Authors in \cite{haase2025sustainability} proposed a task-aware evaluation framework to measure how well LLMs function in workplace settings. Ten practical tasks, such as summarizing texts and writing proposals, were used to evaluate eleven proprietary and open-source LLMs. Despite highlighting sustainability, the study does not directly measure energy consumption and uses indirect factors like the size of the model, the cost per token, and the type of deployment. In contrast, our work focuses on code generation and direct measurements of energy usage, providing an in-depth analysis of LLMs' environmental impact.

Authors in \cite{islam2025evaluating} presented a large-scale study of the energy efficiency of code generated by 20 LLMs across 878 algorithmic programming problems from LeetCode. LLM-generated solutions are compared with human-written code using metrics such as energy and memory consumption. Authors found out that although LLMs produce correct outputs, their code is consistently less energy-efficient than their human counterparts, often by a substantial margin. Authors in \cite{tuttle2024can} evaluated the energy efficiency of code generated by eight state-of-the-art LLMs across eight LeetCode problems using various prompting strategies, and introduced two metrics for comparing LLM-generated code with human-written code, RuntimeRatio and EnergyRatio. Authors in \cite{cursaru2024controlled} evaluated the energy efficiency of Code Llama in comparison to human-written source code. The experiment involves three benchmark tasks implemented in C++, JavaScript, and Python, with Code Llama prompted to generate equivalent implementations using varying prompts and temperature settings. Energy consumption is then measured and compared between the human-generated and LLM-generated versions. Authors in \cite{qiu2024efficient} proposed ENAMEL, a benchmark that measures the efficiency of code generated by LLMs by introducing a novel metric, eff@k, which extends the pass@k metric. 

\begin{figure*}[!t]
    \centering
    \includegraphics[width=0.85\linewidth]{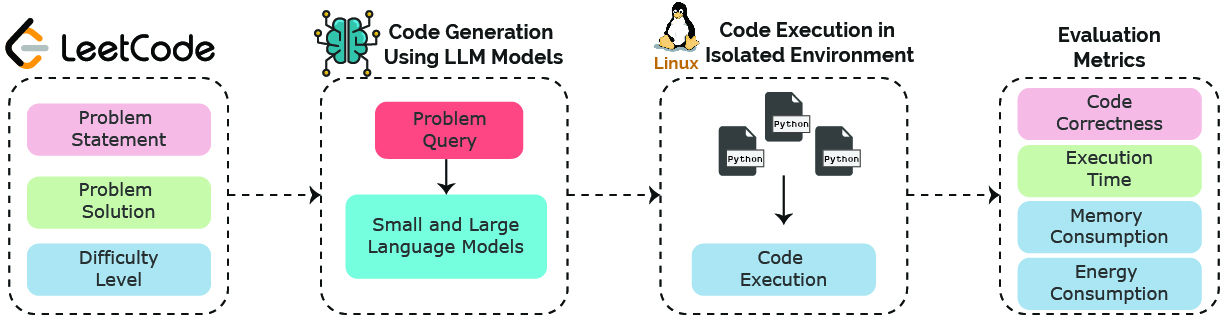}
    \caption{Methodology for Code Generation and Evaluation Using LLM Models}
    \label{fig:model}
\end{figure*}
Authors in \cite{vartziotis2024learn} investigated the energy efficiency of Python code generated by GitHub Copilot, ChatGPT-3, and Amazon CodeWhisperer. According to their findings, AI models can produce more sustainable code when explicitly prompted, but human-written code remains more energy-efficient consistently. As part of AI-assisted software development, authors in \cite{vartziotis2024carbon} examined the carbon footprint of code generated via LLMs within GitHub Copilot and its potential relevance to automotive industries. To evaluate if LLM-generated code adheres to sustainable software engineering principles, they introduce a set of green coding metrics. Nonetheless, the study focuses more on conceptual and qualitative assessments than on empirical evaluations. Finally, authors in \cite{solovyeva2025ai} analyzed the energy efficiency and performance of code generated by LLMs across Python, Java, and C++ on macOS and Windows. A benchmark of  ``hard" programming problems from LeetCode is used to evaluate three advanced LLMs: GitHub Copilot, GPT-4o, and OpenAI o1-mini. The models perform significantly better in generating Python and Java code than C++.

\begin{table}[t!]
\centering
\caption{Example Coding Problems}
\begin{tabular}[t]{cp{0.78\columnwidth}}
\hline
Difficulty & Coding Questions\\
\hline
\hline
Easy & Given an integer $x$, return \texttt{true} if $x$ is a palindrome, and \texttt{false} otherwise.\\
Medium & Given an integer array \texttt{nums} of length $n$, and an integer \texttt{target}, find three integers in \texttt{nums} such that the sum is closest to \texttt{target}.
 \\
Hard & You are given two integer arrays \texttt{nums1} and \texttt{nums2} of lengths $m$ and $n$ respectively. The arrays \texttt{nums1} and \texttt{nums2} represent the digits of two non-negative integers. You are also given an integer $k$. Your task is to create the most significant possible number of length $k$, where $k \leq m + n$, using digits taken from \texttt{nums1} and \texttt{nums2}. The resulting number should preserve the **relative order** of digits taken from the same array.
 \\
\hline
\label{table:questions}
\end{tabular}
\end{table}%
\subsection{Novelty of This Paper}
While previous works \cite{vartziotis2024learn, islam2025evaluating, tuttle2024can, cursaru2024controlled, qiu2024efficient, vartziotis2024carbon, solovyeva2025ai} have explored the energy impact of LLM-generated code, none have explicitly analyzed the energy efficiency of code generation by small LLMs across varying levels of algorithmic complexity. A particular focus of our study is comparing the energy efficiency and performance of code generated by small and LLMs across problems of varying complexity. The goal of this study is to evaluate whether small LLMs are capable of generating sustainable code for particular types of problems and whether they can be applied in AI-assisted software engineering workflows as an alternative to larger, more computationally intensive models. 

This analysis helps researchers and developers understand the trade-offs between model size, energy use, and problem complexity. It supports better decisions when choosing or deploying language models in settings with limited energy and computing resources. Notably, when SLMs demonstrate comparable or superior performance for specific types of programming tasks, they can be effectively deployed in edge devices or local environments. This not only reduces computational overhead but also ensures that sensitive data remains within the local network, enhancing privacy and security without compromising performance.

\section{Methodology}
\label{sec:meth}
The overall methodology of the proposed work is illustrated in Fig. \ref{fig:model}. Following the formulation proposed by Basili et al. \cite{caldiera1994goal}, our high-level goal can be summarized in the following primary research question:

\begin{mdframed}
\textit{How do SLMs and LLMs differ in generating efficient and sustainable code across varying levels of problem complexity?}
\end{mdframed}
Toward this goal, we evaluate and compare the energy consumption and performance metrics of code generated by different language models. The metrics are also compared to human-written solutions that are considered as baselines for efficiency. The primary research question can be broken down into two sub-questions: 

\noindent\textit{\textbf{RQ1:} Can SLMs generate code with comparable performance and efficiency to that of LLMs?} We hypothesize that SLMs can match the performance of LLMs on certain types of programming problems and generate more energy-efficient code, due to inherent differences in their architectures. The goal of this study is to compare the code produced by different models with one another, as well as against baseline human-written solutions.

\noindent\textit{\textbf{RQ2:} How does the energy consumption of code generated by SLMs compare to that of large models and human-written implementations?} In this study, we explore whether small LLMs can generate code that consumes less energy during execution when compared to LLMs and human-written solutions. The main goal is to evaluate the efficiency and sustainability of small LLMs in real-world coding tasks.

\subsection{Selection of Dataset}
We began our study by selecting appropriate coding problems from LeetCode, an educational platform designed to improve programming skills through a variety of coding challenges. In LeetCode, problems are categorized by topic and difficulty level, called \textit{easy}, \textit{medium}, and \textit{hard}. Since Python is widely used across domains and is relevant both in education and industry, we chose to focus on it for our experiments. In total, 150 problems were selected at random, 50 each from easy, medium, and hard categories. We prioritized problems that would allow us to construct appropriate test cases for evaluating the generated code's performance. As part of our evaluation process, we also checked that the chosen problems had community-verified solutions, which we then used as a reference to verify that the outputs generated by the language models were functionally correct. Examples of problems from each difficulty level are presented in Table \ref{table:questions}.

\subsection{Selection of LLMs}
In this study, we used three small LLMs: StableCode-3B\footnote{\url{https://huggingface.co/stabilityai/stable-code-3b}}, trained on over 18 programming languages, and StarCoderBase-3B\footnote{\url{https://huggingface.co/bigcode/starcoderbase-3b}}, trained on over 80 programming languages, and Qwen2.5-Coder-3B-Instruct\footnote{https://huggingface.co/Qwen/Qwen2.5-Coder-3B-Instruct}, trained on 5.5 trillion tokens including source code and text-code grounding data. In addition, we used two LLMs, GPT-4.0 and DeepSeek-Reasoner, and generated code using their respective paid APIs. Small LLMs are selected because they offer the possibility of generating accurate and energy-efficient code with significantly fewer parameters and lower computational requirements. As they are trained in numerous programming languages, they can be applied to a variety of coding tasks, making them ideal candidates for evaluating code sustainability. Furthermore, by including paid LLM models in our evaluation, we aim to provide an accurate comparison that highlights trade-offs between model complexity, performance, and environmental impact.

\subsection{Baseline}

As a baseline, we used human-written solutions for programming problems available on LeetCode. LeetCode has been widely used in research \cite{niu2024evaluating, solovyeva2025ai} because it offers a broad range of problems and ranks solutions based on community votes. This makes it a reliable source for identifying high-quality, efficient code written by experienced developers. For each of the 150 problems in our study, we selected one Python solution that received the highest number of upvotes from the LeetCode community, specifically in terms of clarity, and optimized time and space complexity. These solutions typically include explicit explanations of their computational complexity and are designed to optimize both time and space usage, aligning with LeetCode’s evaluation standards. We use these human-written solutions as a benchmark to compare the efficiency and sustainability of the code generated by both SLMs and LLMs.

\subsection{Sustainability Metrics}
In this study, we use four key metrics to evaluate how efficient and environmentally friendly the generated code is. These metrics are run-time, memory usage, and energy consumption. We also include code correctness to verify that the generated solutions produce the correct results.

\subsubsection{Code Correctness}
We assess the functional correctness of the generated code, as inaccuracies can lead to additional time and computational resources for debugging and fixing, thereby increasing overall resource consumption.

\subsubsection{Run-time}
Run-time refers to the duration the code takes to execute and return a result. It is a critical metric, as longer execution times may reflect inefficiencies in the code's logic or structure. This metric is measured in milliseconds (ms).
\subsubsection{Memory Consumption}
During execution, the code consumes a certain amount of memory, with the highest point being recorded as its peak memory usage. Monitoring this value is essential, as excessive memory consumption can be a barrier to scalability and sustainability, particularly in resource-limited systems. Memory usage is quantified in kibibytes (KiB).
\subsubsection{Energy Consumption}
Energy consumption refers to the total amount of energy used by the code during execution, with a focus on CPU usage. Lower energy usage indicates that the code is more efficient and environmentally sustainable. This metric is measured in  Milliwatt-
hours (mWh).

\subsection{Code Generation} Following the selection of coding tasks and models, the next step was generating code solutions for evaluation. In this study, we investigated three open-source SLMs specifically trained on code datasets, StableCode-3B, StarCoderBase-3B and Qwen2.5-Coder-3B-Instruct, as well as two large commercial models accessed via paid APIs, GPT-4.0 and DeepSeek-Reasoner. Each model received the same input information as a human would when solving a LeetCode problem, including the problem prompt and two test cases with their outputs. This ensured the model knew precisely what the code should do and what the expected result should look like. 

For the open-source models, we downloaded them locally through Hugging Face, then created automated scripts that prompted each model using the original LeetCode problem descriptions. The generated code was then saved as individual Python files for further processing. We developed similar Python scripts to automatically generate responses for the commercial models, which were similarly stored as individual Python files. In cases where code required minor modifications to ensure executability, we used the ChatGPT API to automate the code-cleaning process.

To ensure all generated code could be tested consistently, we provided each model with the same set of instructions. The prompt included the following points:

\begin{itemize}
    \item \textit{Write only valid Python code with no explanations or comments.}
    \item \textit{The code must start with \texttt{class Solution:} and define the solution method inside it.}
    \item \textit{The code must also include an \texttt{if \_\_name\_\_ == "\_\_main\_\_"} block to make it directly runnable.}
\end{itemize}

These instructions helped us run all code samples automatically in a controlled Linux environment. However, in some cases, small models produced more than one solution, even though we asked for only one. When this happened, we used the paid ChatGPT API to clean the code and keep only the first valid solution. This step helped us ensure that all models were compared fairly using one solution per problem. We used Python 3 to both generate and execute the code.

\subsection{Measurement Environment and Experimental Setup} 
The evaluation of sustainability metrics of the generated code was carefully considered, considering both hardware and software factors. All experiments were conducted on Linux in an isolated environment, where Python code generated by SLMs and LLMs was executed and analyzed. To ensure consistency and control, the experiments in this study were conducted on a Google Cloud Compute Engine virtual machine (VM) of type \texttt{c2-standard-8}, located in the \texttt{us-central1-c} region. A C2 VM family offers 8 virtual CPUs supported by Intel 3.9~GHz Cascade Lake processors, along with 32~GB of RAM. It runs Ubuntu 24.04 LTS installed on a 100~GB SSD, and its architecture is \texttt{x86\_64}. A C2 instance was purposefully selected since it offers dedicated CPU cores and a highly stable performance profile, which is essential for collecting reproducible measurements of code execution time, memory footprint, and energy usage. In all cases, Python 3.12.3 was used to execute the scripts. It should be noted that this setup was only used to evaluate the code generated by SLMs and LLMs. To generate the code using small LLMs, we used a Laptop with the following specs for the code generation: CPU:13th Gen Intel(R) Core(TM) i7-13620H   2.40 GHz, RAM: 16 GB DDR5 5200MHz, GPU: Nvidia GeForce RTX 4070 8GB Laptop, and 1TB NVMe SSD.

Each code sample was run ten times to account for variability in performance and to obtain reliable data. To maintain stable conditions, a five-second cooling-down period was applied between executions. For consistency across runs and to minimize the influence of external or nondeterministic factors on the results, all executions were performed under identical virtualized conditions.

\begin{table*}[ht!]
\centering
\caption{Code Analysis Summary by Model and Difficulty Category}
\renewcommand{\arraystretch}{1.2}
\begin{tabular}{llccccccc}
\toprule
\textbf{Model} & \textbf{Category} & \textbf{Total} & \textbf{Logical} & \textbf{Syntax} & \textbf{Correct} & \textbf{Avg Runtime (ms)} & \textbf{Avg Energy (mWh)} & \textbf{Avg Memory (KB)} \\
\midrule
\multirow{3}{*}{StarCoderBase-3B}
 & Easy   & 50 & 17 & 7  & 26 & 22.18 & 1.45 & 629.39 \\
 & Medium & 50 & 28 & 6  & 16 & 22.40 & 1.45 & \textbf{627.62} \\
 & Hard   & 50 & 26 & 11 & 13 & 24.30 & 1.456 & 632.14 \\
\midrule
\multirow{3}{*}{StableCode-3B}
 & Easy   & 50 & 12 & 2  & 36 & 22.45 & 1.45 & 628.32 \\
 & Medium & 50 & 16 & 1  & 33 & 22.95 & 1.452 & 627.89 \\
 & Hard   & 50 & 17 & 5  & 28 & 22.699 & 1.452 & 633.23 \\
\midrule
\multirow{3}{*}{Qwen2.5-Coder-3B-Instruct}
 & Easy   & 50 & 10 & 3  & 37 & 23.99 & 1.458 & \textbf{628.20} \\
 & Medium & 50 & 13 & 1  & 36 & 23.90 & 1.454 & 629.58 \\
 & Hard   & 50 & 15 & 2  & 33 & 24.05 & 1.455 & 634.04 \\
\midrule
\multirow{3}{*}{GPT-4.0}
 & Easy   & 50 & 10 & 0  & 40 & \textbf{19.81} & \textbf{1.444} & 628.62 \\
 & Medium & 50 & 9  & 0  & 41 & \textbf{19.54} & \textbf{1.442} & 630.32 \\
 & Hard   & 50 & 13 & 0  & 37 & \textbf{19.89} & \textbf{1.443} & 633.52 \\
\midrule
\multirow{3}{*}{DeepSeek-Reasoner}
 & Easy   & 50 & 5  & 1  & \textbf{44} & 21.53 & 1.448 & 628.32 \\
 & Medium & 50 & 7  & 0  & \textbf{43} & 22.02 & 1.45 & 627.91 \\
 & Hard   & 50 & 13 & 0  & \textbf{37} & 23.20 & 1.453 & 632.75 \\
\midrule
\multirow{3}{*}{Human-Written}
 & Easy   & \texttt{NA} & \texttt{NA} & \texttt{NA} & \texttt{NA} & 23.56 & 1.453 & 628.34 \\
 & Medium & \texttt{NA} & \texttt{NA} & \texttt{NA} & \texttt{NA} & 23.80 & 1.454 & 629.57 \\
 & Hard   & \texttt{NA} & \texttt{NA} & \texttt{NA} & \texttt{NA} & 24.04 & 1.456 & \textbf{631.6} \\
\bottomrule
\end{tabular}
\label{tab:all_model_analysis_summary}
\end{table*}

\subsection{Analysis of Metrics/ Measurement and Analysis Procedures}

We analyze the generated code samples with a focus on sustainability-related metrics. Specifically, our evaluation includes code correctness, memory usage, energy consumption, and run-time performance.

\subsubsection{Energy Consumption}To estimate the energy consumption of each generated code sample, we use the \texttt{CodeCarbon} Python library. The \texttt{CodeCarbon} tool measures CPU activity to track the energy usage of Python code. 

\subsubsection{Code Correctness}  We developed a script that executed each piece of code using test cases from the LeetCode problem set. The script first checked whether the code ran without errors; if errors occurred, the code was marked as 'no'. Next, it validated the outputs against expected results, marking them as 'yes' if correct and 'no' otherwise. After automated testing was complete, we conducted a manual review of all outputs to identify occasional inaccuracies in automated evaluations, where correct outputs were mistakenly identified as incorrect. 

\subsubsection{Runtime} To measure a code's run-time, we use Python's built-in \texttt{time} module. The \texttt{time} module provides an easy and effective way to track the execution duration of scripts or code blocks. It calculates the total time by recording the wall-clock time before and after execution.

\subsubsection{Memory Usage} To measure the peak memory usage during the execution of each generated code sample, we use Python's built-in \texttt{tracemalloc} module. In each code sample, we activate memory tracing using \texttt{tracemalloc.start()} and retrieve the peak memory usage immediately after execution using \texttt{tracemalloc.get\_traced\_memory()}, which returns both the current and peak memory usage.

\section{Results}
\label{sec:result}
In this section, we present and analyze the evaluation results of various SLMs and LLMs to address our research questions \textbf{RQ1} and \textbf{RQ2}.

\subsection{Code Generation Performance Summary}

Table \ref{tab:all_model_analysis_summary} presents a comprehensive evaluation of SLMs and LLMs along with baseline, across three levels of problem difficulty: Easy, Medium, and Hard. 

In terms of correctness, LLMs still dominate. The DeepSeek-Reasoner solution achieved the highest number of correct solutions across all difficulty levels, 44, 43, and 37 for Easy, Medium, and Hard problems, respectively. GPT-4.0's correct solutions were 40, 41, and 37 for Easy, Medium, and Hard problems. In addition to demonstrating consistent correctness, these models also avoid syntax errors across all levels, indicating strong code structure and language control. Among SLMs, Qwen2.5-Coder-3B-Instruct consistently outperforms StableCode-3B and  StarCoderBase-3B, particularly on harder tasks (Easy: 37, Medium: 36, Hard: 33). On Medium and Hard problems, StarCoderBase-3B shows significantly lower correctness (26–13), compared with StableCode-3B (36–28). Further,  StarCoderBase-3B has the highest number of logical and syntax errors, with 28 logical and 6 syntax errors on Medium problems, indicating its low reasoning capabilities. Based on these results, Qwen2.5-Coder-3B-Instruct is highly competitive with larger models in terms of correctness, and it exhibits strong generalization across all problem difficulties.

Across all difficulty levels, GPT-4.0 has the fastest execution time, with average runtimes of 19.81 ms (Easy), 19.54 ms (Medium), and 19.89 ms (Hard). Compared to other models, these are significantly faster than Qwen and StableCode-3B, which consistently exceed 22 ms. For instance, Qwen2.5-Coder-3B-Instruct runs at 23.90 ms to 24.00 ms, which is the slowest overall. Despite Qwen2.5-Coder-3B-Instruct's competitive performance in correctness, the generated code may involve more complex logic or structural overhead, which results in longer execution times. These results highlight that LLMs such as GPT-4.0 and DeepSeek-Reasoner are significantly more efficient in terms of runtime, consistently outperforming SLMs and human-written code across all difficulty levels.

In terms of energy consumption, across all difficulty levels, GPT-4.0 consistently shows the lowest energy usage, ranging from 1.442 mWh (Medium) to 1.444 mWh (Easy), making it the most energy-efficient model. The consumption level of DeepSeek-Reasoner follows close behind, at 1.448–1.453 mWh. As for the smaller models,  StarCoderBase-3B, StableCode-3B, and Qwen2.5-Coder-3B-Instruct show slightly higher energy usage, typically around 1.45–1.458 mWh, with Qwen2.5-Coder-3B-Instruct showing the highest energy usage. They suggest that the more accurate outputs from smaller models like Qwen2.5-Coder-3B-Instruct come at a modest cost in energy, although the differences are minor.

\subsection{Success Rate Comparison Across LLMs and SLMs}
\begin{figure}[t!]
    \centering
    \includegraphics[width=\linewidth]{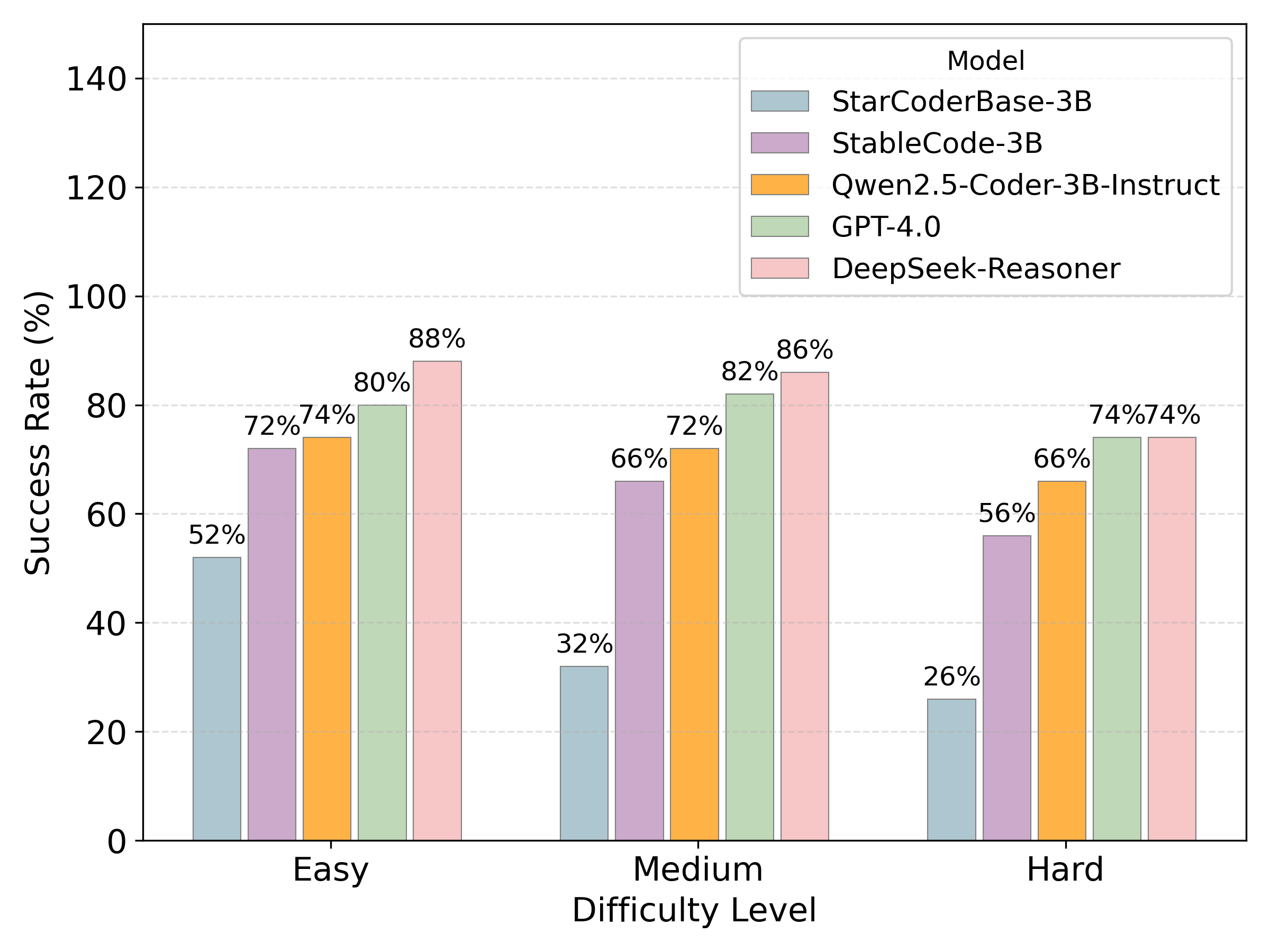}
    \caption{Success Rate Comparison of SLMs and LLMs}
    \label{fig:success_rate_comparison}
\end{figure}

Fig. \ref{fig:success_rate_comparison} shows a comparison of the success rates across three difficulty levels, defined as the percentage of output codes that are correct out of 150 attempts. Among Easy, Medium, and Hard problems, DeepSeek-Reasoner achieves the highest success rate, with 88\%, 86\%, and 74\% success rates. GPT-4.0 follows closely with success rates of 80\%, 82\%, and 74\%. Qwen2.5-Coder-3B-Instruct performs best among the SLMs with 74\% on Easy, 72\% on Medium, and 66\% on Hard problems, substantially narrowing the performance gap with LLMs. Secondly, StableCode performs moderately, while  StarCoderBase-3B consistently underperforms, achieving only 26\% success on hard tasks. Despite LLMs' superior accuracy and generalization across varying levels of problem complexity, advanced SLMs such as Qwen2.5-Coder-3B-Instruct can still offer competitive performance, especially in resource-constrained settings.

\begin{tcolorbox}[colback=gray!10, colframe=black, boxrule=0.3mm, arc=0mm, left=2mm, right=2mm, top=1mm, bottom=1mm]
\textbf{Summary:} The LLMs achieved the highest level of correctness in all difficulty levels and also performed better on both runtime and energy efficiency than SLMs. There is a general trend for SLMs to perform less accurately than LLMs, though Qwen2.5-Coder-3B-Instruct exhibited excellent generalization and competitive accuracy. Commercial LLMs continue to face challenges in achieving perfect accuracy, primarily due to persistent problems in code generation and occasional hallucinations..
\end{tcolorbox}

\subsection{Energy Efficiency in Correct Outputs by SLMs}
Fig. \ref{fig:slm_energy_efficiency_analysis} illustrates the number of problems for which SLMs, specifically Qwen2.5-Coder-3B-Instruct,  StarCoderBase-3B, and StableCode-3B, produced accurate outputs and consumed the same amount of energy or less than LLMs, namely GPT-4.0 and DeepSeek-Reasoner, along with human-written solutions. Results are categorized by problem difficulty: Easy, Medium, and Hard, along with a Total category that aggregates all levels. According to the analysis, SLMs generated correct outputs and also achieved better or equal energy efficiency for 33 Easy problems, accounting for 66\% of the set. For Medium and Hard problems, this occurred for 21 (42\%) and 25 (50\%) problems, respectively. A total of 79 out of 150 problems were found to produce the correct solution while matching or outperforming the LLMs in terms of energy efficiency. Although the correctness accuracy of LLMs was significantly higher than that of SLMs, analysis reveals that when SLMs did produce correct outputs, those solutions were often energy-efficient as well.

\begin{tcolorbox}[colback=gray!10, colframe=black, boxrule=0.3mm, arc=0mm, left=2mm, right=2mm, top=1mm, bottom=1mm]
\textbf{Summary:} While SLMs generally exhibit lower overall correctness compared to LLMs, they achieve notable energy efficiency when their outputs are correct. In 52.6\% of the total problems, at least one SLM produced a correct and energy-efficient solution comparable to LLMs.
\end{tcolorbox}

\begin{table}[ht]
\centering
\caption{Energy-Efficient Outputs by SLMs When Code is Correct (Count and Percentage per Difficulty Level, out of 50 problems)}
\begin{tabular}{|l|c|c|c|}
\hline
\textbf{Difficulty} & \textbf{Qwen2.5-Coder} & \textbf{StableCode} & \textbf{ StarCoderBase} \\
\hline
Easy   & 13 (26\%) & 11 (22\%) & 9 (18\%)  \\
Medium &  9 (18\%) &  8 (16\%) & 4 (8\%)   \\
Hard   & 11 (22\%) & 10 (20\%) & 4 (8\%)   \\
\hline
\textbf{Overall} & \textbf{33 (22\%)} & \textbf{29 (19.3\%)} & \textbf{17 (11.3\%)} \\
\hline
\end{tabular}
\label{tab:slm_correct_combined}
\end{table}
\subsection{Performance Comparison of Correct Outputs by SLMs}
\begin{figure}[t!]
    \centering
    \includegraphics[width=0.96\linewidth]{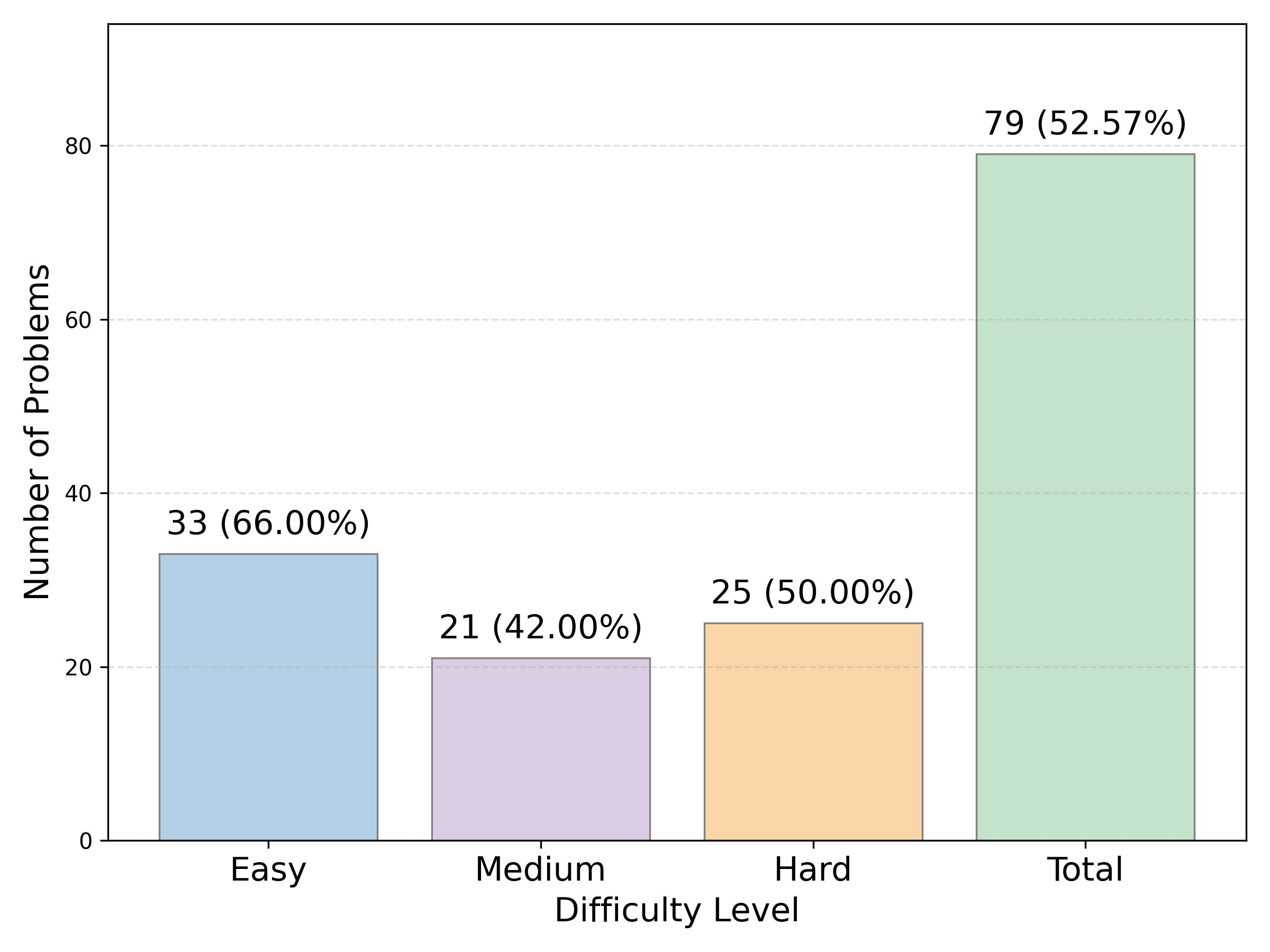}
    \caption{Energy-efficient correctness comparison across difficulty levels, showing how often SLMs generated correct code with equal or lower energy consumption than LLMs or human baselines. The Total bar aggregates all 150 problems.}
    \label{fig:slm_energy_efficiency_analysis}
\end{figure}

To further break down the analysis presented in Fig. \ref{fig:slm_energy_efficiency_analysis}, Table \ref{tab:slm_correct_combined} presents the number and percentage of correct outputs achieved by each SLM across three difficulty categories. Each percentage is calculated out of 50 problems per category, allowing us to assess the relative performance of each SLM in terms of sustainability metrics. Among SLMs, Qwen2.5-Coder-3B-Instruct performs best, with 13 correct solutions (26\%) in Easy problems, 9 (18\%) in Medium problems, and 11 (22\%) in Hard problems, resulting in 33 correct outputs (22\%). Based on Easy and Medium outputs, StableCode-3B scores 11 correctly (22\%), 8 correctly (16\%), and 10 correctly (20\%), making 29 in total (19.3\%). With just 9 (18\%), 4 (\%), and 4 (8\%) correct outputs for Easy, Medium, and Hard, respectively,  StarCoderBase-3B shows the lowest correctness. The results demonstrate Qwen2.5-Coder-3B-Instruct's consistency in generalizing across difficulty levels, making it the most accurate SLM in the study. StableCode-3B performs with moderate accuracy, while StarCoderBase-3B shows the lowest performance, especially on harder problems, suggesting limited reasoning ability.

\begin{tcolorbox}[colback=gray!10, colframe=black, boxrule=0.3mm, arc=0mm, left=2mm, right=2mm, top=1mm, bottom=1mm]
\textbf{Summary:} Among all SLMs, Qwen2.5-Coder-3B-Instruct demonstrates the highest accuracy and the most consistent performance across different problems. Results also show that not all SLMs perform equally well, so choosing the right model is important, even when energy usage is similar.
\end{tcolorbox}

\section{Limitations}
\label{sec:ttv}
A few limitations should be acknowledged in this study when comparing SLMs and LLMs in Python-based code generation. We first evaluate three small open-source LLMs and two large commercial LLMs accessed via APIs. Although these models represent current capabilities, the findings may not apply to all LLMs, especially those using newer architectures or trained on different datasets. Additionally, LLMs are inherently nondeterministic, and the same prompt may yield different results across multiple runs, introducing variability. There is also a possibility that some benchmark coding problems could have been incorporated into the training data of certain LLMs, resulting in memorization and overestimation of performance.

Secondly, the study only addresses Python, a widely used language, which may not reflect the characteristics of other languages such as C++ and Java. Therefore, our conclusions are not generalizable across programming paradigms. Additionally, we use human-written solutions voted on by the community on LeetCode as a baseline for performance. Even though these are typically high-quality posts, upvotes may not always reflect optimal efficiency. Lastly, although we conducted experiments in a controlled and isolated Linux environment, minor system fluctuations and measurement overhead may result in slight errors in runtime, energy, or memory measurements.

\section{Conclusion}
\label{sec:con}
This study compares SLMs and LLMs for automated code generation, focusing on their performance, energy efficiency, and correctness across varying Python problems. According to our results, while LLMs such as GPT-4.0 and DeepSeek-Reasoner consistently achieve higher correctness rates and faster runtimes, SLMs offer noticeable advantages in energy efficiency, especially when their outputs are accurate. In comparison with other SLMs evaluated, Qwen2.5-Coder has the highest accuracy and best generalization across the three difficulty levels, outperforming all other SLMs. The fact that SLMs consume the same or less energy as LLMs in more than half of the problems where they produced correct outputs further supports their potential for deployment in resource-constrained environments. Furthermore, the results demonstrate that not all SLMs perform equally well, highlighting the importance of model selection, even when energy consumption appears similar. The analysis will help researchers and developers better understand the trade-offs between model size, energy efficiency, and problem complexity, enabling more informed decisions when selecting or deploying LLMs for code generation in energy-constrained environments.

\bibliographystyle{IEEEtran}
\bibliography{biliography}
\end{document}